\documentstyle[preprint]{ptptex}

\newcommand{\bra}[1]{\langle {#1} |}     
\newcommand{\ket}[1]{| {#1} \rangle}     
\newcommand{\wtilde}[1]{\widetilde{#1}} 

\title{
Note on the Deformed Boson Scheme in\\
Time-Dependent Variational Method
}

\author{
Atsushi {\sc Kuriyama}, 
Constan\c{c}a {\sc Provid\^encia}$^{*}$, \\
Jo\~ao da {\sc Provid\^encia}$^{*}$, Yasuhiko {\sc Tsue}$^{**}$ 
and Masatoshi {\sc Yamamura}
}

\inst{
Faculty of Engineering, Kansai University, Suita 564-8680, Japan\\
$^{*}$Departamento de Fisica, Universidade de Coimbra, P-3000 Coimbra, 
Portugal\\
$^{**}$Physics Division, Faculty of Science, Kochi University, Kochi 780-8520, 
Japan
}

\recdate{
}

\abst{
The Holstein-Primakoff representation for the $su(2)$-algebra is 
derived 
in the deformed boson scheme. 
The following two points are discussed : (i) connection 
between a simple Hamiltonian and the Hamiltonian obeying the 
$su(2)$-algebra 
such as Lipkin model and (ii) 
derivation of the Hamiltonian for describing the damped and amplified 
motion for the $su(2)$-boson model. 
}

\begin{document}

\maketitle


The $q$-algebra has been receiving much attention in physics. 
If the generators of the algebra are expressed in terms of 
boson operator, for example, such as the Schwinger boson representation 
for the $su(2)$-algebra,\cite{Schw} the $q$-deformed algebra is 
constructed in the framework of the deformed boson scheme including 
conventional $q$-deformation, 
which is in close contact with generalized boson coherent state 
for the time-dependent variational method. 
In this scheme, the function $[x]_q$ plays a central role. 
Concerning the choice of $[x]_q$, Penson and Solomon proposed 
an interesting viewpoint.\cite{Penson} 
By generalizing their viewpoint, the present authors, recently, 
proposed a possible method for the deformed boson scheme 
in the time-dependent variational method (Part (I))\cite{KPPTYI} 
and, further, they applied this method to some algebras 
(Parts (II) and (III)).\cite{KPPTYII}
Hereafter, Part (I) is referred to as (I). However, in (I), a certain 
case is excluded. In this note, this case is investigated.

First, we recapitulate a possible form of the deformed boson 
scheme presented in (I). The basic point of (I) is to define the 
deformed boson annihilation operator ${\hat \gamma}$, 
the eigenstate of which is the following generalized coherent 
state for boson operator $({\hat c} , {\hat c}^*)$ : 
\begin{eqnarray}
& &\ket{c}=\left(\sqrt{\Gamma}\right)^{-1}
\sum_{n=0}^{\infty} f(n)(\sqrt{n!})^{-1}\gamma^n \ket{n} \ ,
\label{L1} \\
& &\ket{n}=(\sqrt{n!})^{-1} \left({\hat c}^*\right)^n\ket{0} \ , 
\nonumber\\
& &[ {\hat c} , {\hat c}^*]=1 \ , \qquad 
{\hat c}\ket{0}=0 \ .
\label{L2}
\end{eqnarray}
Here, $\Gamma$ and $\gamma$ denote the normalization factor and 
a complex parameter, respectively. 
For the function $f(n)$, in (I), we treated the case 
\begin{equation}\label{L3}
f(n)=1 \quad \hbox{\rm for}\quad n=0, 1 \ , \qquad f(n)>0 \quad \hbox{\rm for}
\quad n=2, 3, \cdots \ .
\end{equation}
The state (\ref{L1}) can be rewritten in the following form : 
\begin{eqnarray}
& &\ket{c}=\left(\sqrt{\Gamma}\right)^{-1}
\exp \left(\gamma{\hat c}^* {\wtilde f}({\hat N})\right) \ket{0} \ ,
\label{L4}\\
& &{\hat N}={\hat c}^*{\hat c} \ . 
\label{L5}
\end{eqnarray}
Here, ${\wtilde f}(n)=f(n+1)f(n)^{-1}$. 
The condition (\ref{L3}) is equivalent to 
\begin{equation}\label{L6}
{\wtilde f}(n)=1 \quad \hbox{\rm for} \quad n=0 \ , \qquad
{\wtilde f}(n) > 0 \quad \hbox{\rm for}\quad n=1, 2, 3, \cdots \ .
\end{equation}
We introduce the operators ${\hat \gamma}$ and ${\hat \gamma}^*$ in 
the form 
\begin{equation}\label{L7}
{\hat \gamma}={\wtilde f}({\hat N})^{-1}{\hat c} \ , \qquad 
{\hat \gamma}^*={\hat c}^* {\wtilde f}({\hat N})^{-1} \ .
\end{equation}
It is easily shown that the operator ${\hat \gamma}$ satisfies the relation 
\begin{equation}\label{L8}
{\hat \gamma}\ket{c}=\gamma\ket{c} \ .
\end{equation}
The above means that $({\hat \gamma} , {\hat \gamma}^*)$ can be 
regarded as the deformed boson operator of $({\hat c} , {\hat c}^*)$. 
The commutation relation $[{\hat \gamma} , {\hat \gamma}^* ]$, 
which is deformed from $[{\hat c} , {\hat c}^*]=1$, is rewritten as 
\begin{equation}\label{L9}
[{\hat \gamma} , {\hat \gamma}^*] = [{\hat N}+1]_q - [{\hat N}]_q \ , \qquad
[{\hat N}]_q={\hat N}{\wtilde f}({\hat N}-1)^{-2} \ .
\end{equation}
The above is an outline of (I).

In this note, we give a possible form of $({\hat \gamma} , {\hat \gamma}^*)$ 
obeying the following condition : 
\begin{eqnarray}\label{L10}
& &{\wtilde f}(n)=1 \quad \hbox{\rm for}\quad n=0 \ , \qquad
{\wtilde f}(n)>0 \quad \hbox{\rm for}\quad n=1, 2, \cdots , n_0-1 \ ,
\nonumber\\
& &{\wtilde f}(n) \longrightarrow \infty \quad \hbox{\rm for}\quad
n=n_0 \ , \nonumber\\
& &{\wtilde f}(n) : \hbox{\rm arbitrary} \quad \hbox{\rm for}\quad 
n=n_0+1, n_0+2, \cdots \ .
\end{eqnarray}
Later, the example will be shown. For the condition (\ref{L10}), 
we define the state 
\begin{equation}\label{L11}
\ket{c}=\left(\sqrt{\Gamma}\right)^{-1}
\exp\left(\gamma{\hat c}^*{\wtilde f}({\hat N}){\hat P}_{n_0-1}\right)
\ket{0} \ .
\end{equation}
Here, ${\hat P}_{n_0-1}$ denotes a projection operator obtained by 
replacing $n$ with $n_0-1$ for 
\begin{equation}\label{L12}
{\hat P}_n=\sum_{k=0}^n \ket{k}\bra{k} 
=1-\left[{\hat c}^*\left(\sqrt{{\hat N}+1}\right)^{-1}\right]^{n+1}
\left[\left(\sqrt{{\hat N}+1}\right)^{-1}{\hat c}\right]^{n+1} \ .
\end{equation}
The state (\ref{L11}) is rewritten as 
\begin{subequations}\label{L13}
\begin{eqnarray}
& &\ket{c}=\left(\sqrt{\Gamma}\right)^{-1}
\left( \ket{0}+\sum_{n=1}^{n_0}({\wtilde f}(0){\wtilde f}(1)\cdots
{\wtilde f}(n-1))(\sqrt{n!})^{-1}\gamma^n \ket{n}\right) \ , 
\label{L13a}\\
& &\Gamma=1+\sum_{n=1}^{n_0}(|\gamma|^2)^n
({\wtilde f}(0){\wtilde f}(1)\cdots
{\wtilde f}(n-1))^2(n!)^{-1} \ . 
\label{L13b}
\end{eqnarray}
\end{subequations}
For the state (\ref{L11}), we define ${\hat \gamma}$ and 
${\hat \gamma}^*$ in the form 
\begin{equation}\label{L14}
{\hat \gamma}={\hat P}_{n_0}{\wtilde f}({\hat N})^{-1}{\hat c} \ , 
\qquad
{\hat \gamma}^*={\hat c}^*{\wtilde f}({\hat N})^{-1}{\hat P}_{n_0} \ .
\end{equation}
Here, ${\hat P}_{n_0}$ is obtained by replacing $n$ with $n_0$ 
for ${\hat P}_n$ shown in the relation (\ref{L12}). 
The operation of ${\hat \gamma}$ on the state ({\ref{L11}) gives us 
\begin{equation}\label{L15}
{\hat \gamma}\ket{c}=\gamma{\hat P}_{n_0-1}\ket{c} \ .
\end{equation}
Further, we have 
\begin{equation}\label{L16}
{\hat \gamma}\ket{n_0+1}=0 \ , \qquad
{\hat \gamma}^*\ket{n_0}=0 \ .
\end{equation}
Therefore, the subspace spanned by the set $\{ \ket{n} ; n=0, 1, 
\cdots, n_0\}$ is disconnected with the subspace $\{ \ket{n} ; 
n=n_0+1, n_0+2 , \cdots \}$ through ${\hat \gamma}$ and ${\hat \gamma}^*$. 
The relation (\ref{L15}) shows that $\ket{c}$ is not an exact eigenstate 
for ${\hat \gamma}$. However, at the limit $n_0\rightarrow \infty$, 
the relation (\ref{L15}) reduces to the relation (\ref{L8}). 
In this sense, we regard $({\hat \gamma} , {\hat \gamma}^*)$ given 
in the relation (\ref{L15}) as a possible form of the deformed 
boson obeying the condition (\ref{L10}). 
With use of the relation $[{\hat P}_{n_0} , {\wtilde f}({\hat N})^{-1}
{\hat c} ] = [{\hat P}_{n_0} , {\hat c}^*{\wtilde f}({\hat N})^{-1} ]=0$, 
we have 
\begin{equation}\label{L17}
[{\hat \gamma} , {\hat \gamma}^* ] 
={\hat P}_{n_0}\left([{\hat N}+1]_q - [{\hat N}]_q\right){\hat P}_{n_0} \ , 
\qquad [{\hat N}]_q={\hat N}{\wtilde f}({\hat N}-1)^{-2} \ .
\end{equation}
The above is a possible form of the deformed boson scheme 
obeying the condition (\ref{L10}).

Now, let us present a concrete example : 
\begin{equation}\label{L18}
{\wtilde f}(n)=\left( \sqrt{1-n/n_0}\right)^{-1} \ .
\end{equation}
Certainly, the form (\ref{L18}) satisfies the condition (\ref{L10}). 
Then, ${\hat \gamma}$ and ${\hat \gamma}^*$ can be expressed as 
\begin{eqnarray}\label{L19}
& &{\hat \gamma}={\hat P}_{n_0}\sqrt{1-{\hat N}/n_0}\ {\hat c}
{\hat P}_{n_0}=(\sqrt{n_0})^{-1}\cdot{\hat {\cal S}}_- \ , \nonumber\\
& &{\hat \gamma}^*={\hat P}_{n_0}{\hat c}^*
\sqrt{1-{\hat N}/n_0}\ 
{\hat P}_{n_0}=(\sqrt{n_0})^{-1}\cdot{\hat {\cal S}}_+ \ .
\end{eqnarray}
Here, ${\hat {\cal S}}_\pm$ are defined in the form 
\begin{subequations}\label{L20}
\begin{eqnarray}
& &{\hat{\cal S}}_\pm={\hat P}_{n_0}{\hat S}_\pm {\hat P}_{n_0} \ , 
\label{L20a}\\
& &{\hat S}_-=\sqrt{n_0-{\hat N}}\ {\hat c} \ , \qquad
{\hat S}_+={\hat c}^*\sqrt{n_0-{\hat N}} \ .
\label{L20b}
\end{eqnarray}
\end{subequations}
Further, we have 
\begin{equation}\label{L21}
[{\hat \gamma} , {\hat \gamma}^* ]={\hat P}_{n_0}
(1-2{\hat N}/n_0){\hat P}_{n_0} = -2n_0^{-1}{\hat {\cal S}}_0 \ , 
\end{equation}
\vspace{-0.8cm}
\begin{subequations}\label{L22}
\begin{eqnarray}
& &{\hat {\cal S}}_0={\hat P}_{n_0}{\hat S}_0{\hat P}_{n_0} \ , 
\label{L22a}\\
& &{\hat S}_0={\hat N}-n_0/2 \ .
\label{L22b}
\end{eqnarray}
\end{subequations}
We can see that the form ${\hat S}_{\pm, 0}$ given in the relations 
(\ref{L20b}) and (\ref{L22b}) are identical to the Holstein-Primakoff 
representation of the $su(2)$-spin, the magnitude of which is 
equal to $n_0/2$.\cite{HP}
Importance of the operator ${\hat P}_{n_0}$ in the form 
${\hat {\cal S}}_{\pm, 0}={\hat P}_{n_0}{\hat S}_{\pm, 0}{\hat P}_{n_0}$ 
was demonstrated by Marshalek\cite{M} and reinvestigated by the 
present authors.\cite{KPTY7}

In relation to the above investigation, the $su(1,1)$-spin in the 
Holstein-Primakoff representation is also derived. In this case, 
we adopt the following form as ${\wtilde f}(n)$ : 
\begin{equation}\label{L23}
{\wtilde f}(n)=\left(\sqrt{1+n/n_0}\right)^{-1} \ .
\end{equation}
Here, it should be noted that $n_0$ is positive, not necessary 
positive integer and ${\wtilde f}(n)$ obeys the 
condition (\ref{L6}). Then, applying the form (\ref{L23}) to 
$({\hat \gamma} , {\hat \gamma}^*)$ defined in the relation (\ref{L7}), 
we have 
\begin{eqnarray}\label{L24}
& &{\hat \gamma}=(\sqrt{n_0})^{-1}{\hat T}_- \ , \qquad
{\hat T}_-=\sqrt{n_0+{\hat N}}\ {\hat c} \ , \nonumber\\
& &{\hat \gamma}^*=(\sqrt{n_0})^{-1}{\hat T}_+ \ , \qquad
{\hat T}_+={\hat c}^*\sqrt{n_0+{\hat N}} \ , \nonumber\\
& &[{\hat \gamma} , {\hat \gamma}^* ]=2n_0^{-1}{\hat T}_0 \ , \qquad
{\hat T}_0={\hat N}+n_0/2 \ . 
\end{eqnarray}
The form of ${\hat T}_{\pm, 0}$ is identical to the Holstein-Primakoff 
representation of the $su(1,1)$-spin, the magnitude of which is 
equal to $n_0/2$.

Let us apply the above deformed boson scheme to the following 
Hamiltonian : 
\begin{subequations}\label{L25}
\begin{eqnarray}
& &{\hat H}={\hat H}_0+{\hat H}_1 \ , 
\label{L25o}\\
& &\ {\hat H}_0=2\epsilon\ {\hat c}^*{\hat c}
=2\epsilon\ {\hat c}^* \cdot 2(1+[{\hat c} , {\hat c}^*])^{-1}\cdot
{\hat c} \ , 
\label{L25a}\\
& &\ {\hat H}_1=-G_0{\hat c}^*{\hat c}-(G_1/2)({\hat c}^{*2}+{\hat c}^2) 
\ . \label{L25b}
\end{eqnarray}
\end{subequations}
The deformed Hamiltonian ${\hat{\cal H}}$ can be given in the form 
\begin{subequations}\label{L26}
\begin{eqnarray}
& &{\hat {\cal H}}={\hat{\cal H}}_0+{\hat{\cal H}}_1 \ , 
\label{L26o}\\
& &\ {\hat {\cal H}}_0=2\epsilon{\hat \gamma}^* 
\cdot 2(1+[{\hat \gamma} , {\hat \gamma}^*])^{-1}\cdot
{\hat \gamma} \ , 
\label{L26a}\\
& &\ {\hat {\cal H}}_1=-G_0{\hat \gamma}^*{\hat \gamma}
-(G_1/2)({\hat \gamma}^{*2}+{\hat \gamma}^2) 
\ . \label{L26b}
\end{eqnarray}
\end{subequations}
For the function (\ref{L18}), ${\hat{\cal H}}_0$ and ${\hat{\cal H}}_1$ 
can be expressed in the form 
\begin{subequations}\label{L27}
\begin{eqnarray}
& &{\hat{\cal H}}_0={\hat P}_{n_0}
\left[2\epsilon (n_0/2+{\hat S}_0)\right]{\hat P}_{n_0} \ , 
\label{L27a}\\
& &{\hat {\cal H}}_1={\hat P}_{n_0}
\left[-(G_0/n_0){\hat S}_+{\hat S}_-
-(G_1/2n_0)({\hat S}_+^2+{\hat S}_-^2)\right]{\hat P}_{n_0} \ .
\label{L27b}
\end{eqnarray}
\end{subequations}
The above form is identical to the Hamiltonian obeying the 
$su(2)$-algebra, for example, such as Lipkin model.\cite{LMG}
Of course, for ${\hat S}_{\pm, 0}$, the forms (\ref{L20b}) and 
(\ref{L22b}) should be used.\cite{YK,KM} 
It may be interesting to see that the deformed boson scheme 
for simple boson Hamiltonian gives us the Hamiltonian obeying 
the $su(2)$-algebra. However, it may be noted that the derivation is in 
a trick. Since $[ {\hat c} , {\hat c}^* ]=1$, the Hamiltonian (\ref{L25}) 
can be expressed as a function of $[ {\hat c} , {\hat c}^*]$ in 
infinite ways. Of course, they are all equivalent to one another. 
However, if $[{\hat c} , {\hat c}^*]$ is replaced with 
$[{\hat \gamma} , {\hat \gamma}^*]$, they are not equivalent to 
one another. Therefore, the above is nothing but one of the examples for the 
deformation. It is also possible to derive the other various forms, 
if we choose ${\wtilde f}(n)$ in forms different from the 
form (\ref{L18}).

Finally, we discuss a many-boson system consisting of two kinds of 
boson operators $({\hat b}, {\hat b}^*)$ and $({\hat a} , {\hat a}^*)$. 
The typical example for such systems may be given by the following 
Hamiltonian : 
\begin{subequations}\label{L28}
\begin{eqnarray}
& &{\hat H}={\hat H}_b-{\hat H}_a-{\hat H}_i \ , 
\label{L28o}\\
& &\ {\hat H}_b=\omega{\hat b}^*{\hat b} \ , \qquad
{\hat H}_a=\omega{\hat a}^*{\hat a} \ , 
\label{L28a}\\
& &\ {\hat H}_i=\gamma\cdot i({\hat b}^*{\hat a}^*-{\hat a}{\hat b}) \ .
\label{L28b}
\end{eqnarray}
\end{subequations}
Here, $\omega$ and $\gamma$ denote positive parameters. The above 
Hamiltonian was investigated by Vitiello et al.\cite{CRTV} and 
slightly later by the present authors.\cite{TKYI} 
It enables us to give a possible description of the damped- and 
the amplified-oscillation in the framework of the conservative form. 
We generalize the Hamiltonian (\ref{L28a}) to the form 
\begin{eqnarray}\label{L29}
& &{\hat H}_b=2\epsilon {\hat b}^*{\hat b}-G_0{\hat b}^*{\hat b}
-(G_1/2)({\hat b}^{*2}+{\hat b}^2) \ , \nonumber\\
& &{\hat H}_a=2\epsilon{\hat a}^*{\hat a}-G_0{\hat a}^*{\hat a}
-(G_1/2)({\hat a}^{*2}+{\hat a}^2) \ .
\end{eqnarray}
In the same idea as that used for the Hamiltonian (\ref{L25}), 
we have the following Hamiltonian : 
\begin{subequations}\label{L30}
\begin{eqnarray}
& &{\hat{\cal H}}={\hat {\cal H}}_b-{\hat{\cal H}}_a-{\hat{\cal H}}_i \ , 
\label{L30o}\\
& &{\hat{\cal H}}_b={\hat P}_{n_0}^{(b)}
\bigl[2\epsilon(n_0/2+{\hat S}_0^{(b)})\nonumber\\
& &\qquad\qquad\quad
-(G_0/n_0){\hat S}_+^{(b)}{\hat S}_-^{(b)}-(G_1/2n_0)
({\hat S}_+^{(b)2}+{\hat S}_-^{(b)2})\bigl]
{\hat P}_{n_0}^{(b)} \ , \nonumber\\
& &{\hat{\cal H}}_a={\hat P}_{n_0}^{(a)}
\bigl[2\epsilon(n_0/2+{\hat S}_0^{(a)})\nonumber\\
& &\qquad\qquad\quad
-(G_0/n_0){\hat S}_+^{(a)}{\hat S}_-^{(a)}-(G_1/2n_0)
({\hat S}_+^{(a)2}+{\hat S}_-^{(a)2})\bigl]
{\hat P}_{n_0}^{(a)} \ , 
\label{L30a}\\
& &{\hat{\cal H}}_i={\hat P}_{n_0}^{(b)}{\hat P}_{n_0}^{(a)}
\left[(\gamma/n_0)\cdot i
\left({\hat S}_+^{(b)}{\hat S}_+^{(a)}-{\hat S}_-^{(a)}{\hat S}_-^{(b)}
\right)\right]{\hat P}_{n_0}^{(a)}{\hat P}_{n_0}^{(b)} \ .
\label{L30b}
\end{eqnarray}
\end{subequations}
Here, the quantities with the indices $b$ and $a$ are related to 
${\hat b}$- and ${\hat a}$-bosons, respectively. The Hamiltonian 
(\ref{L30}) enables us to describe the damped and the amplified 
motion in the $su(2)$-spin system, which was already investigated by 
the present authors.\cite{TKYII}
Especially, the form of ${\hat{\cal H}}_i$ was already introduced 
in a form different from the present case (\ref{L30b}), but, 
essentially the same.

Thus, we could present the case which was excluded in (I). 
The essential difference from (I) is in the existence of the 
projection operator ${\hat P}_{n_0}$.


\end{document}